# Hot Carrier Dynamics in Operational Metal Halide Perovskite Solar Cells

Hadi Afshari[1], Varun Mapara[1], Shashi Sourabh[1], Megh N. Khanal[2], Vincent R. Whiteside[2], Rebecca A. Scheidt[3], Matthew C. Beard[3], Giles E. Eperon[4], Ian R. Sellers[2], and Madalina Furis[1]

[1] *Department of Physics & Astronomy, University of Oklahoma, Norman, OK 73019, U.S.A*

[2] *Department of Electrical Engineering, University at Buffalo, NY 14260, U.S.A*

[3] *National Renewable Energy Laboratory, Golden, CO 80401, U.S.A*

[4] *Swift Solar, San Carlos, Ca, 94070, U.S.A*

**Abstract**

One of the main approaches to inhibit carrier cooling in semiconductor systems enabling the study of hot carrier solar cell protocols is the use of concentrated illumination to obtain high power densities and create a phonon bottleneck. This, however, typically also increases the lattice temperature of the solar cells significantly. Accordingly, the solar cells subject to high concentration illumination also need to withstand high operating temperatures. Having previously demonstrated the high temperature tolerance of the triple halide perovskite ($FA_{0.8}Cs_{0.2}Pb_{1.02}I_{2.4}Br_{0.6}Cl_{0.02}$) solar cells, here the hot carrier relaxation dynamics are studied in these devices using high power transient absorption (TA) measurements. In addition to monitoring TA spectra obtained at different time delays, the thermalization mechanisms of hot carriers is mapped with power dependent TA to extract the carrier cooling time in this system under in-operando conditions at various bias conditions that reflect the $J_{sc}$, $V_{max}$ and $V_{oc}$ of these structures, and subsequently deconvolve the underlying physics of carrier relaxation; as well as track the dynamics of the thermalization close to working conditions of the solar cells. These measurements uncover a complex interaction of hot carrier thermalization involving the temporal carrier density, transport, and extraction, and apparent non-equivalent contributions with respect to non-equilibrium photogenerated electrons and holes in these metal halide perovskite solar cell architectures.

**Introduction**

Historically the application of solar cell technology started in space to power satellites [1, 2]. Today however, solar panels have become ubiquitous and in many different forms, finding applications beyond conventional roof top and utility solar installations to agrivoltaics, where the dual utilization of land for agriculture and solar farming proves lucrative for investors while conserving water through substantial reductions in irrigation [3, 4]. As well as a new generation of electric vehicles that incorporate solar roofs, capable of generating extra driving miles exclusively from sunlight, particularly advantageous in sunny cities [5-7]. Despite these advancements, the full potential of solar cell technology is yet to be realized. The primary work to maximize the utilization of the sun's energy lies in addressing the largest loss mechanisms within solar cells, notably thermalization and transmission losses [8, 9]. Thermalization occurs when high-energy photons are absorbed by the solar cell, creating high energy carriers that thermalize to the band edges [8]. This process results in the biggest loss in solar cells, approximately 35% of the solar spectrum is wasted due to thermalization, and the parasitic heat generation that results [8, 10]. The second significant loss in solar cells is transmission loss, whereby photons with energies lower than the band gap pass through the device [8]. Both of these loss mechanisms can be somewhat reduced by using multijunction solar cell technology. However, the cost of this technology is currently prohibitive and limits its use to only space photovoltaics (PV) [11, 12].

The hot carrier solar cell (HCSC) has long been considered as a potential protocol to circumvent thermalization losses in single band gap solar cells and aims to harvest high energy electrons (holes) prior to excess energy dissipation [13, 14]. Specifically, in semiconductors, excitation above the band gap results in high energy "hot" carriers (HCs) with a distribution temperature $T_c$ (that forms via carrier-carrier interactions) far exceeding that of the lattice [15-17]. In most cases the hot carrier population reaches a quasi-equilibrium state in a sub-picosecond time scale. Subsequent, slower carrier cooling ensues through predominately carrier-phonon scattering, which is dominated by the emission of LO phonons in polar semiconductors – such as the metal halide perovskites studied here – until thermal equilibrium between the carriers and the lattice is reached [13, 16].

Hot carrier solar cell architectures aim to increase the intra-band relaxation time of the photogenerated carriers to values close to the carrier extraction time, such that carriers retain their kinetic energy upon collection [18]. Significantly, if achieved, such hot carrier solar cells promise power conversion efficiencies in excess of 60 % for single junction solar cells [14, 15].

At present, there are two primary approaches towards hot carrier solar cells that aim at slowing down non-equilibrium carrier cooling: (1) engineered quantum confinement in systems such as nanocrystals or multiple quantum well structures and; (2) creating a phonon-bottleneck in systems through high excitation densities using concentrated photoexcitation [19-21]. In both scenarios, the cooling time may be extended to tens of picoseconds, influenced by the synergistic effects of the intrinsic phonon-bottleneck, intensified Auger heating effects, and recently the effects of valley scattering [13, 22-24].

While considerable work has been performed in the traditional III-V materials [25], recently there has also been growing interest in the metal halide perovskites as a potential system for the realization of the HCSC with many recent works suggesting the presence of long-lived hot carriers in these systems [21, 26-29]. Although there is some debate in the community regarding the carrier thermalization pathways and the competing thermalization mechanisms [15, 27, 30-32]. Recently there has been evidence of hot carriers in perovskite device structures [21, 27, 29] providing encouragement for the potential of metal halide perovskites in hot carrier device applications. Much of the early research in this field focused on hot carrier dynamics using ultrafast spectroscopy, while solar cell devices operate in the CW or steady-state regime. Key innovations in the new cell architecture, (which enabled high power pulsed and steady-state assessment of perovskites), have boosted their stability under both elevated temperatures and high-fluence photoexcitation, which are necessary to stimulate non-destructive hot carrier populations [17, 27, 33].

While CW spectroscopy and time resolved techniques have traditionally been used to assess hot carriers in optical structures, here hot carrier effects and dynamics in metal halide perovskite solar cells in the ultrafast regime are assessed *in operando*. Specifically, the cooling dynamics of the hot carriers in triple halide perovskite ($FA_{0.8}Cs_{0.2}Pb_{1.02}I_{2.4}Br_{0.6}Cl_{0.02}$) solar cells are studied using power dependent (PD) transient absorption (TA) measurements under various biasing conditions including: (1) at $V_{oc}$ open circuit with no current extraction (2) at $V_{Jsc} = 0.02$ V, close to $J_{sc}$ with maximum current passing through the solar cell and at (3) $V_{max}$ or the maximum power point of these devices. These data provide interesting insight into hot carrier dynamics in solar cells under practical operating conditions and the role that carrier absorption and extraction play in the formation and dissipation of hot carrier distributions in these systems. All these measurements were therefore performed using both front and back illumination, the orientation of which is described more fully below.

**Materials and Methods**

The perovskite precursor chemicals were utilized in their original form and stored in a nitrogen glovebox. The preparation of perovskite solutions and the deposition of the film were carried out in the same nitrogen glovebox. Solutions for wide-gap perovskite precursors were created by dissolving Formamidinium Iodide (Greatcell), Cesium Iodide (Sigma Aldrich), Lead (II) Iodide (TCI), Lead (II) bromide, and Lead (II) Chloride (Alfa Aesar). This resulted in a 1.2 M solution of $FA_{0.8}Cs_{0.2}PbI_{2.4}Br_{0.6}Cl_{0.02}$ in DMF:DMSO, with a volumetric ratio of 3:1. It is important to note that the inclusion of a small amount of Cl is intentionally in stoichiometric excess [33].

ITO-coated glass substrates from Thin Film Devices underwent an initial cleaning process involving sonication in acetone and propan-2-ol, followed by UV ozone treatment for 10 minutes. PolyTPD (Poly(N,N'-bis-4-butylphenyl-N,N'-bisphenyl)benzidine) was applied through spin-coating from a 1 mg/mL solution in anhydrous chlorobenzene at 4000 rpm for 30 seconds. The solution was deposited dynamically and subsequently annealed at 110 °C for 10 minutes [33].

PFN-Br (poly[(9,9-bis(3′-(N,N-dimethylamino)propyl)-2,7-fluorene)-alt-2,7-(9,9-dioctylfluorene)]) was utilized as a wetting layer, spin-coated from a 0.5 mg/mL solution in anhydrous methanol at 5000 rpm for 20 seconds. The solution was deposited dynamically. The perovskite film was then spin-coated at 5000 rpm for 60 seconds, with the solution spread on the substrate before spinning. Additionally, 120 μL of anhydrous methyl acetate was deposited onto the spinning wet film at 25 seconds into the spin cycle. The resulting film underwent annealing at 120 °C for 20 minutes [33].

Following perovskite film deposition, a 1 nm layer of lithium fluoride was thermally evaporated at a rate of 0.1 Å/s. Subsequently, 30 nm of $C_{60}$ (Lumtec) was thermally evaporated at a rate of 0.2 Å/s for the first 10 nm and 0.5 Å/s for the remaining 20 nm. A thin ALD nucleation layer of PEIE (polyethylenimine ethoxylated) was deposited through spin-coating a 0.025 wt % solution, prepared by diluting the purchased PEIE/water solution (40 wt %, Sigma-Aldrich), with anhydrous propan-2-ol. This layer was spin-coated at 5000 rpm for 20 seconds and deposited statically before spinning. The films were then annealed at 100 °C for 2 minutes.

Subsequently, 20 nm of $SnO_x$ was deposited via ALD (Beneq TFS200 ALD), followed by 2 nm of zinc tin oxide (ZTO). An ITO electrode was deposited through room-temperature sputtering using a shadow mask (300 nm) in a Denton Explorer sputter tool. Finally, an oxide encapsulant layer was deposited via ALD, comprising 50 nm of $Al_2O_3$ interspersed with cycles of $TiO_2$.

The perovskite solar cells are measured using various characterizing techniques. The photovoltaic characteristics are examined through *J-V* measurements under the AM 1.5G solar spectrum, employing the Newport Oriel Sol2A solar simulator. A Keithley 2400 Source meter is employed to apply bias to the test cells and simultaneously measure the extracted current from the device [33]. Pump-probe transient absorption (TA) measurements were carried out at room temperature and various electrical bias conditions using a HELIOS TA spectrometer manufactured by Ultrafast Systems. The measurement employed a pump beam at 442 nm 2 kHz of an Apollo-Y Optical Parametric Amplifier (OPA). The white light probe was generated in a sapphire crystal located inside the HELIOS spectrometer using the focused 1064 nm 2 kHz output of a Hyperion femtosecond amplified laser. The pulse duration of both pump and probe are approximately 350 fs.

**Results and Discussions**

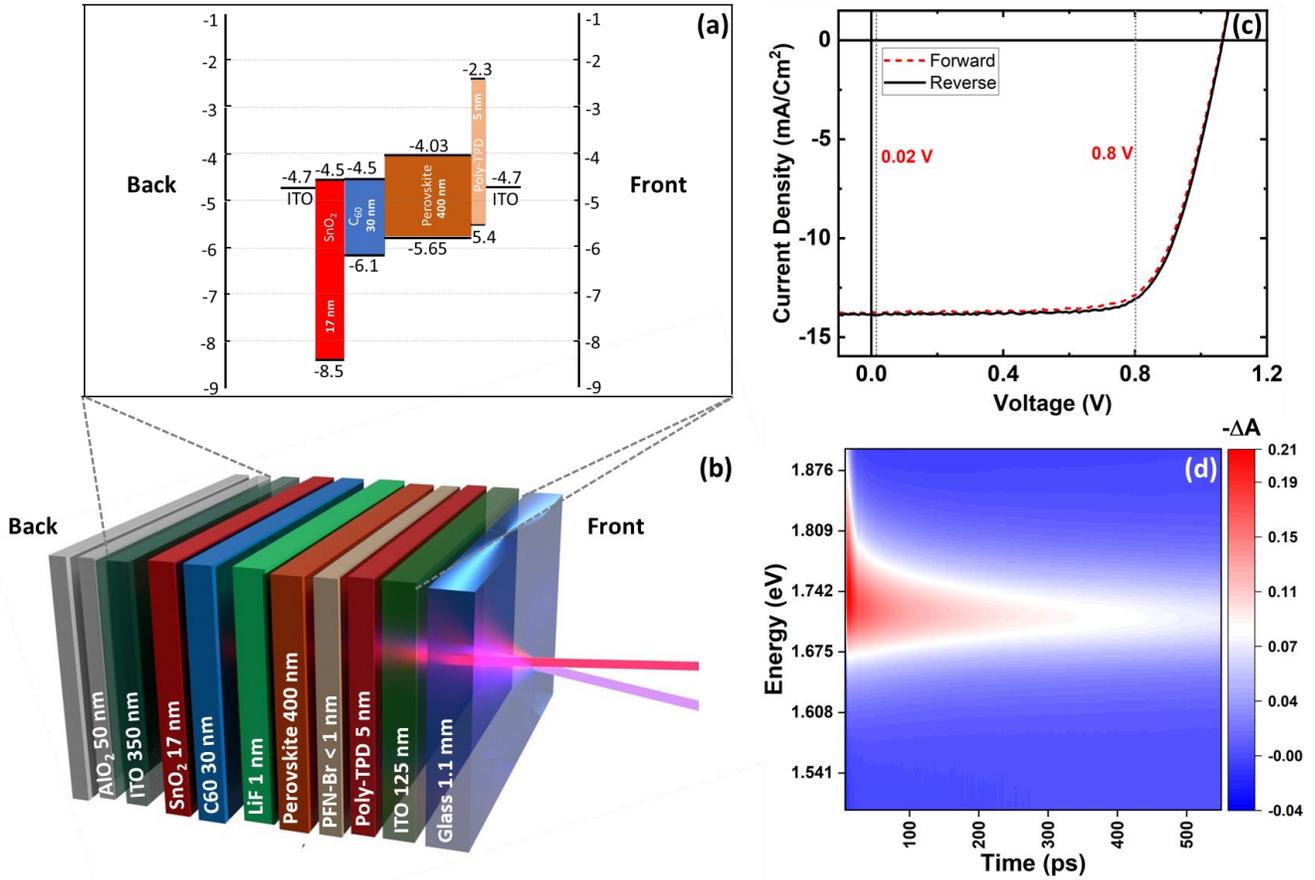

**Figure 1(a). Schematic illustration of the band diagram of the metal halide perovskite solar cell under investigation. (b) Illustrative depiction of structure of the solar cell indicating the front and back side orientation. (c) A comparison of the current density – voltage (J-V) response of the solar cell measured forward (dashed red line) and reverse (solid black line) direction under 1-sun AM 1.5G. The vertical dashed lines indicate the voltages applied for the various TA measurements presented below. (d) Representative heat map of the transient (TA) at intermediate power at 300 K. The blue color indicates the small change in the absorption (-$\Delta A$) and red indicates the higher change in absorption (-$\Delta A$).**

Figure 1(a) shows a schematic of the energy band diagram of the solar cell assessed in this work. This architecture follows a typical perovskite solar cell comprising a heterostructure device which consists of a Poly-TPD hole transport layer (HTL) and $C_{60}$ and $SnO_2$ electron transport layers (ETL). These encapsulate a 400 nm halide perovskite absorber layer, and the structure is completed with both upper and lower ITO layers to enable good device contacting and subsequent front and back transmission measurements. The "front side" of the device is the side through which solar irradiation typically impinges in these devices through the glass substrate and subsequent hole transport layer (see Figure 1(a)). Based on this notation, the "back" of the device therefore describes an orientation in which excitation occurs through the ITO covered $SnO_2/C_{60}$ ETL. Figure 1(b) shows the full stack structure of the solar cell and illustrates the direction of the co-incident and overlapping pump (442 nm) and probe (white light visible range) beams under front side excitation (approx. 50 µm in diameter). The $k$-vector of the incident probe beam is perpendicular to the device layers as also shown in Figure 1(b).

Figure 1(c) shows the light $J$-$V$ graph of the solar cell taken under 1-sun AM1.5 conditions. The $J$-$V$ measurement is performed in forward and reverse with negligible evidence of hysteresis demonstrating the high quality of the device. While the $J_{sc}$ of ~ 13 mA/cm$^2$ reflects the non-optimum absorber thickness fabricated to facilitate absorption measurements, the quality of this device is evident by the high $V_{oc}$ of ~ 1.1 V measured for these structures. The $J$-$V$ response shown in Figure 1(c) also indicates as vertical *dashed lines* the biases at which the power dependent TA (PD-TA) measurements were taken on this device. An additional TA measurement was also performed where the solar cell was not electrically contacted (*true $V_{oc}$*), henceforth this condition is referred to as the "open circuit".

A typical and representative TA heatmap (a color 2D plot of ΔA vs energy and time in pico-seconds) is shown as an example in Figure 1(d). This TA response exhibits the typical bleaching feature associated with inter band absorption at the band gap energy with a high energy tail at early delay times attributed to the presence of a high energy hot carrier population, enabling the extraction and investigation of hot carrier dynamics in these systems [15, 21, 34].

In order to fully interpret the hot carrier dynamics in TA measurements the role of recombination and its effect on temporal carrier population is critical. The upper panel of Figure 2 describes the evolution of the non-equilibrium carrier distribution modeled using a standard Fermi gas formalism, and the mechanisms that dissipate excess energy following the photoexcitation of carrier on a timescale spanning from femtoseconds to one second. Following photoexcitation, within a matter of several hundred femtoseconds, the initial carrier distribution, dictated by the energy and intensity of the pump, promptly undergoes elastic scattering via carrier-carrier interactions. This process leads to the establishment of a Fermi–Dirac distribution characterized by an extremely high non-equilibrium carrier temperature. Over the subsequent tens of picoseconds, the carriers experience inelastic scattering, effectively thermalizing the distribution.

The lower panel of Figure 2 illustrates the effect of thermalization on the carrier distribution (calculated for the system studied here), over almost 2 orders of magnitude in temporal window from sub *ps* to tens of *ps*. Within sub-picosecond regime (red region), elastic carrier-carrier scattering occurs, facilitating the redistribution of energy among photogenerated carriers as shown in the distribution labeled 0.4 *ps*. The photogenerated electrons (holes) rapidly attain a global non-equilibrium carrier distribution within the conduction (valence) band, known as hot carrier distribution, accompanied by a carrier temperature exceeding the lattice's equilibrium temperature ($t < 2\ ps$). However, within a few tens of picoseconds post-light absorption (amber region, picoseconds range), carrier-phonon interactions dominate, leading to gradual thermalization of the carriers and the subsequent dissipation of excess energy and heat from the system in the 500 *ps* range.

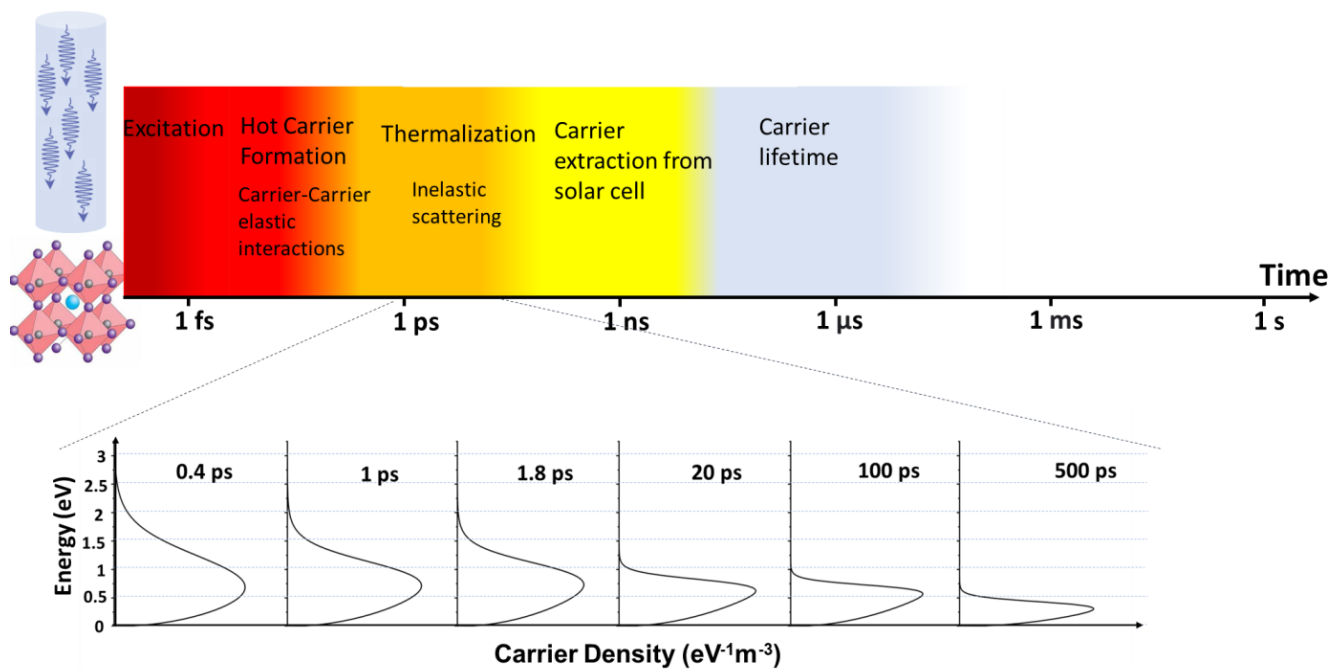

**Figure 2. The upper panel illustrates the sequence of processes of a semiconductor following photo-excitation, as depicted in the order shown. These include hot carrier formation within tens of femtoseconds, hot carrier thermalization over tens of picoseconds, carrier extraction from the solar cell occurring within a few to tens of nanoseconds, and carrier lifetimes that can extend up to microseconds. The lower panel focuses on the thermalization phase, showing the calculated carrier distribution for the studied system at various delay times. At 0.4 picoseconds, the data reveals the highest population of hot carriers, with longer delay times corresponding to a reduced population of these carriers.**

Figure 3(a) shows the evolution of the TA spectra with increasing pump-probe delay times, for the case of the non-contacted solar cell illuminated from front side (the results for other biases and conditions are presented in the supplementary information, Figure S1). Spectra recorded at early delay times exhibit a high energy tail associated with distribution of hot carriers [35-37]. At long delay times the photobleach experiences a redshift and linewidth ($\Gamma$) narrowing associated with hot carrier thermalization and carriers cooling to the band edge [37, 38]. It is noteworthy to mention that in some materials the photobleach behavior is opposite in terms of full width at half maximum (FWHM), reporting broadening of $\Gamma$ for longer delay times [21]. Upon photoexcitation the Burstein–Moss effect, (which reflects state filling due to the high rate of carrier generation) competes with band gap renormalization, (that is associated with coulombic repulsion due to the high density of photogenerated carriers), which induces a competing redshift.

In addition to the dynamics related to the band gap bleach, photo-induced absorption (PIA) is also evident in the TA at both sides of the main photobleach peak Figure 3(a). While below band gap PIA is explained through (probe-induced) excitation of photogenerated (pump-induced) carriers, the above-band gap PIA has also previously been ascribed to the change in the imaginary part of the refractive index in metal halide perovskites, which serves to modulate the optical response of these materials [15, 38, 39].

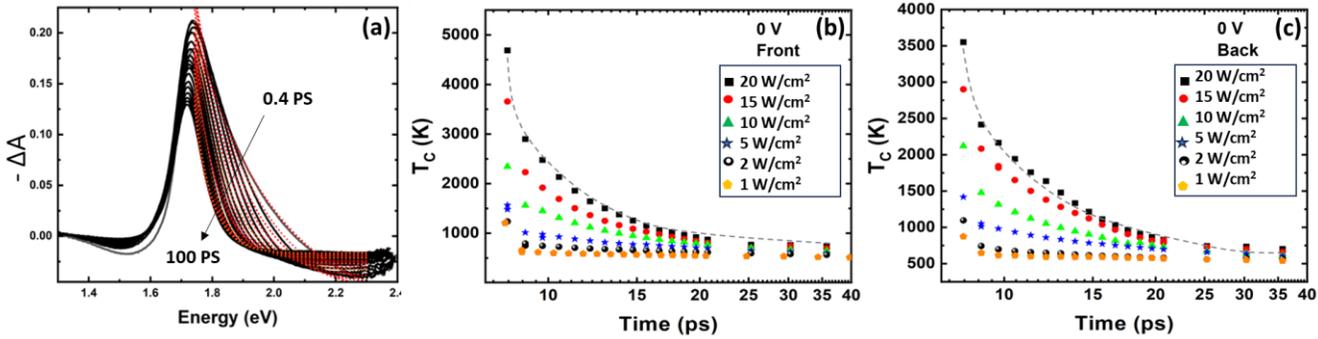

**Figure 3. (a) Transient absorption of the solar cell excited from the front side of the structure. The red dotted lines represent the fitting of the high energy side of the curves with a generalized form of Planck's radiation law, Equation (1). The extracted carrier temperatures, $T_c$, for front (b) and back side (c) illumination, respectively.**

On inspection the TA shown in Figure 3(a) shows that up to ~ 6 *ps*, the magnitude of photobleaching increases, while the FWHM of the TA narrows (TA curves focused on this time window are depicted in Supplementary Information, Figure S2). This pattern qualitatively follows the Fermi-Dirac distribution of the electrons in the conduction band (holes in the valence band) illustrated in Figure 2 after excitation, which is explained through relaxation of hot carriers to lower energy levels in the conduction band. At increasing delay times, $t > 6\ ps$, the magnitude of the bleach systematically decreases due to a combination of radiative recombination and lateral redistribution of the carriers towards regions outside of the probe beam zone. Also, the presence of built-in fields at both interfaces (HTL and ETL) of the absorber layer, which extend towards the center of the metal halide perovskite layer cause drifting of carriers towards the transport layers and therefore reduce the carrier density ($n_0$) in the absorber, which can lower the magnitude of the bleach. The reduction in the carrier density is accounted for when plotting the hot carrier distribution at longer time delays up to 500 *ps* in the lower panel of Figure 2.

As seen in Figure 3(a) while the low energy side of the TA curves are relatively unchanged (except for the reduction of the PIA at longer delay times) with time, the high energy side changes dramatically. These changes reflect the temporal evolution and thermalization of the photogenerated hot carrier distribution. The carrier temperature, $T_c$, can be extracted from the high energy side of the TA curves using the Fermi-Dirac distribution function, which for hot carrier conditions $((E - E_F) \gg k_B T_c)$, is approximated with Maxwell–Boltzmann distribution, yielding [15, 21]:

$$-\Delta A(\hbar\omega) = -A_0(\hbar\omega) \exp\left(-\frac{\hbar\omega - E_F}{k_B T_c}\right) \quad (1)$$

Or alternatively one can use the relation:

$$\frac{\Delta T}{T}(E) \propto \exp\left(-\frac{E - E_F}{k_B T_c}\right) \quad (2)$$

where $E_F$ is the quasi-Fermi level, $k_B$ is Boltzmann constant, and $T_c$ is hot carrier temperature. This is a technique developed by the HC community to evaluate HC dynamics and temperature across several systems [15, 21, 40-42]. Here, simple "tail fitting" is used to gain a *qualitative* assessment of high energy carrier relaxation/dynamics as well as, the role of bias and/or excitation geometry. Example of the "fits" used to determine $T_c$ from Equation (2) are shown as red dotted lines in Figure 3(a) for the device under non-contact bias-free frontside illumination.

Figures 3(b) and (c) show the laser fluence (power) dependence of $T_c$ extracted from a non-contacted ($V_{oc}$) device illuminated through the frontside glass substrate, see Figure 1(a), and also back side illumination (the supplementary Information, Figure S3 has the same data for cases $V_{Jsc}$ and $V_{max}$). Within the femtosecond time frame (red section) after post-photoexcitation, the photogenerated carriers form a high temperature hot carrier distribution. Over tens *to* hundreds of femtoseconds elastic carrier-carrier interactions dominate resulting in non-equilibrium carrier temperatures, $T_c$, as high as 5000 K, in this case. As shown in Figure(s) 3(b) and (c) higher $T_c$ are achieved for higher power densities. For example, at 1 W/cm$^2$ (10 Suns) $T_c$ reaches ~ 1000 K while for the 20 W/cm$^2$ (200 Suns) the temperature is > 4000 K.

The power independent regimes evident in these data at $t > 10\ ps – 20\ ps$ are considered the equilibrium regime, where the carriers are thermalized and are nearly at the lattice temperature.

When considering the direction of excitation, illumination through the glass substrate and subsequent HTLs (denoted "front") results in higher $T_c$, as compared to the illumination through the "back" (bc) and ETLs. Specifically, $T_c$ at 0.4 $ps$ and for 20 W/cm$^2$ is ~ 4700 K for frontside (fr) excitation, and ~ 3600 K for backside illumination. Although the origin of the non-uniform $T_c$ with respect to excitation direction requires further investigation, the non-uniform carrier temperatures suggest a subtle interplay between carrier absorption and extraction/transfer in these architectures. Under frontside illumination more photons are absorbed within the active perovskite layer (with respect to backside illumination through the $C_{60}$/$SnO_2$) due to the relatively thin wide gap poly-TPD layer that comprises the HTL, and that absorbs little of the incident light under direct excitation (see Supplemental Figure S4). The larger relative absorption under frontside illumination increases the generated carrier density in the MHP layer and subsequently the relative phonon bottleneck that would be generated.

As is illustrated in Figure 1(a), incident light at the back side initially traverses and is absorbed in a relatively thicker electron transport layer comprising $C_{60}$ and $SnO_2$. $C_{60}$ in particular, has a high absorption coefficient in the visible range (~1.8 eV, direct band gap), which will inevitably result in approximately 20 % less of the 442 nm photons from the pump beam reaching the halide perovskite layer under back side excitation, when compared to illumination of the device through the front side (thinner poly-TPD). The relative absorption through these two geometries is illustrated in the supplementary Figure S4, which shows a transfer matrix simulation of the structures under front and back side illumination, with the lower photon transmission at the back. It is therefore postulated that the difference in $T_c$ observed in Figure(s) 3(b) and (c) reflects the difference in photogenerated carriers absorbed in the perovskite film under the two excitation conditions, which is lower under back side illumination (with respect to the front) reducing the relative phonon bottleneck – which is highly carrier density dependent – and therefore $T_c$ induced under excitation from the back (relative to front excitation).

However, here it must also be noted, that the $T_c$ extracted from the TA reflects the average temperature of the carriers, comprising both hot electron *and* hot hole thermalization, as such little is known *directly* regarding the relative temperature and thermalization of these constituent carriers. Additionally postulated, is the relative carrier transport and extraction efficiency affects the "average" carrier temperature and that depending upon excitation orientation the respective thermalization rate of electrons and holes is different. This is discussed further below.

In Figure 3(b) and (c) – irrespective of the relative carrier temperature, $T_c$ systematically reduces in both excitation orientations to a temperature approaching that of the lattice (equilibrium), a process mediated by carrier-carrier and carrier-phonon scattering such that the photogenerated carrier population equilibrates with the thermal reservoir.

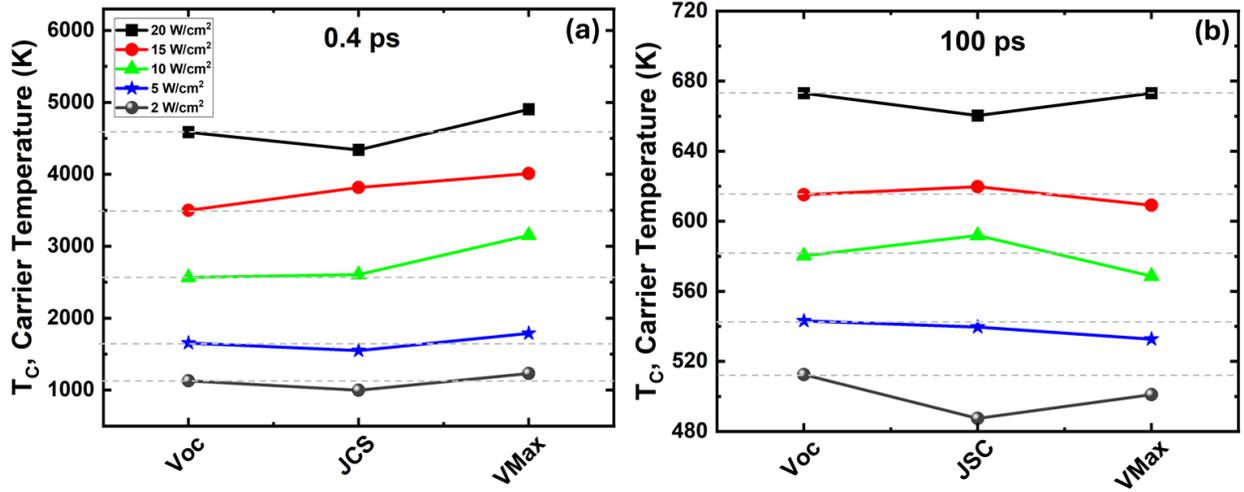

**Figure 4. (a) Carrier temperatures at 0.4 *ps* for all the laser pump powers as a function of external bias. Illumination is incident from the frontside (b) Carrier temperatures at 100 *ps* for all the powers of the frontside illumination as a function of external bias. Illumination is incident from the frontside.**

Figure 4 compares the carrier temperatures at (a) earliest delay time, 0.4 *ps* (b) and after thermalization is mostly complete, 100 *ps*, for different powers and biases ((1) $V_{oc}$, not contacted, (2) $V_{Jsc}$ = 0.02 V, and (3) $V_{max}$ = 0.8 V) under frontside illumination. Interestingly, in Figure 4(a) in the initial fast decay regime (0.4 *ps*) there appears a power dependent increase in the carrier thermalization at $V_{max}$. In this temporal regime, carrier cooling is dominated by Frohlich interactions and cooling via carrier-LO phonon interactions [10, 15, 43]. At $V_{max}$, the carrier transport through the absorber is dominated by diffuse currents that are strongly coupled to the lattice via the polar nature of these systems. Indeed, polarons are known to contribute considerably to the optoelectronic properties of metal halide perovskites [44-46] and even suggested to result in hot carrier stabilization via hot phonon reabsorption [37], which may reflect the inhibited thermalization observed at $V_{max}$ under high hot carrier generation induced at higher power in Figure 4(a).

For the 100 *ps* case the carriers have the lowest temperature at $V_{max}$, see Figure 4(b), except for the highest power of 20 W/cm$^2$. At $V_{max}$ the built-in field is actively exerting force on photogenerated carriers and there is $I_{max}$ current through the solar cell. Apparently at the earliest times, the combination of non-zero internal electric field and current transfer initially impacts the carrier distribution promoting carrier-carrier interactions that lead to higher carrier temperatures. This is in accordance with the BGR effect (intensified under $V_{max}$ conditions) in halide perovskites which dominates at early time delays before the hot carriers start to thermalize. At sufficiently longer times, the presence of the built-in field and $I_{max}$ - in the case of $V_{max}$ - does not apparently facilitate the stabilization of hot carriers causing equilibrate faster. In Figure 4(b) there is one exception, at ~ 20 W/cm$^2$, in which the $T_c$ at $V_{max}$ is equal to the $T_c$ at $V_{oc}$, which may reflect the very high carrier temperature at these conditions at initial (0.4 *ps*) times, see Figure 4(a).

To further evaluate the hot carrier dynamics, $T_c$, was also extracted using Equation (1) under the various photovoltaic operational regimes (or applied biases). In addition to those data presented in Figure 3(b) and (c) at (1) $V_{oc}$, $T_c$ was also extracted under bias at (2) 0.02 V ~ $V_{Jsc}$, and (3) 0.8 V ~ $V_{max}$, (see Figure S3) as deduced from the *J-V* response under ambient conditions at 1-sun AM 1.5G illumination (see Figure 1(c)).

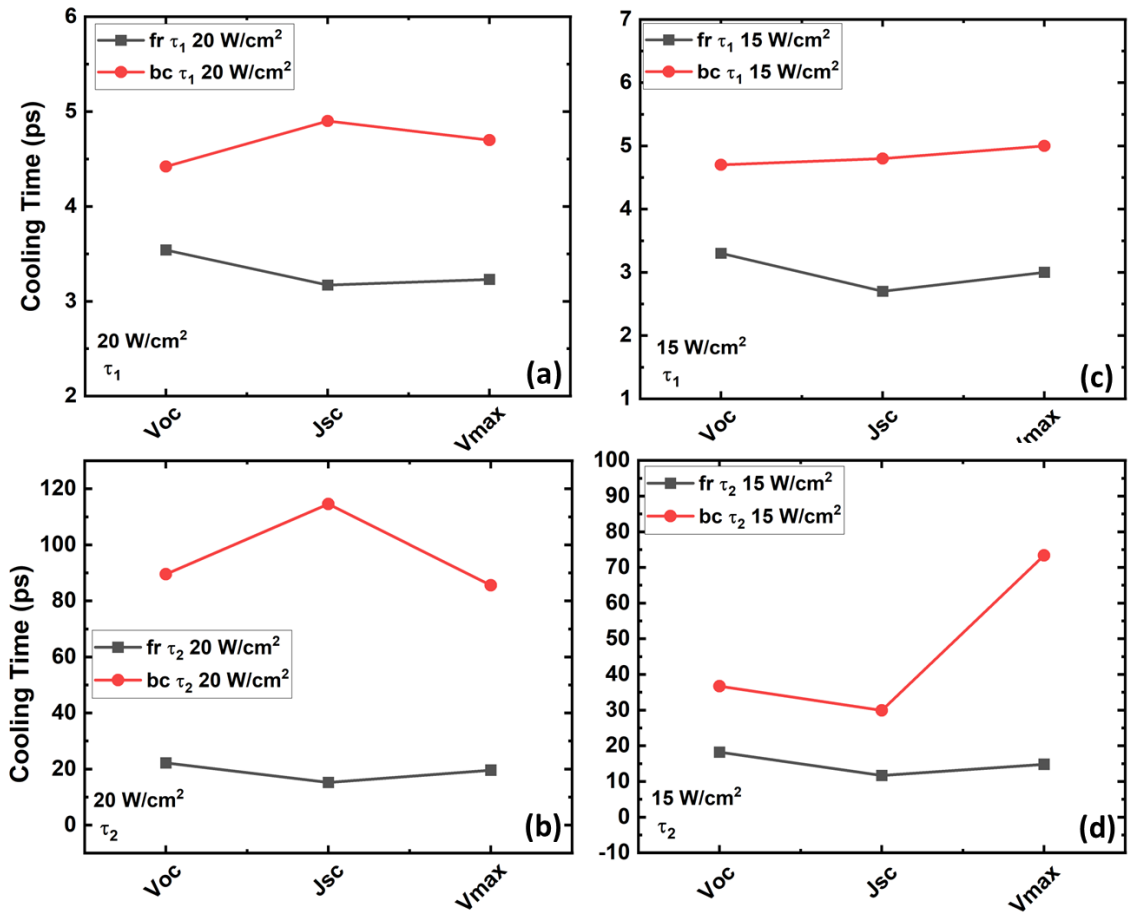

**Figure 5.** Extracted cooling times at power densities of 15 W/cm² and 20 W/cm² for both front and back side illumination, measured under various bias conditions. A short cooling time ($\tau_1$) in the range of 3–5 *ps* and a longer cooling time ($\tau_2$) in the range of 10–120 *ps* are extracted from the fitting

of carrier temperature vs. time responses. Figures (a) and (b) display the results for $\tau_1$ and $\tau_2$ at 20 W/cm², while figures (c) and (d) present the results for $\tau_1$ and $\tau_2$ at 15 W/cm². In all cases, the black squares represent front side illumination results, and the red circles represent back side illumination results.

The carrier cooling time (thermalization rate) is extracted from full biexponential fits to the temporal evolution of carrier temperature responses (see Figures 3(b) and (c)) with the extracted parameters $\tau_1$ and $\tau_2$ describing intra-band relaxation – thermalization – of the carriers at open circuit and the dissipation of excess heat in the system, respectively; for data under the other various biases see the supplementary information (Figure S3 ). A comparison of cooling times $\tau_1$ and $\tau_2$ - or the temporal decays – at various bias conditions at higher excitation powers, are presented in Figure 5. It is observed that for higher powers, namely 1200 µW (15 W/cm²) and 1600 µW (20 W/cm²), $\tau_1$ and $\tau_2$ have a bias dependence, while for the low powers this behavior is not apparent. Figures 5(a) and (c) show the relative cooling time at shorter times ($t < 5\ ps$) at 15 W/cm² and 20 W/cm², respectively, while Figure(s) 5(b) and (d) show the behavior at $t > \sim 10\ ps$, also for the respective two powers discussed. The fast and slower heat dissipation regimes observed in (for example) Figure 3 have been reported previously in the metal halide perovskites and have been attributed to a combination of phonon bottleneck effects, and Auger re-heating effects, at moderate to high carrier densities [15, 21, 34, 35]. More recently, it has also been suggested that the slower cooling regime is observed across the metal halide perovskites is due to simple lattice heating and the low thermal conductivity of the metal halide perovskites, in general [27, 47]; that is, only the fast component of the decay curves relates to the hot carrier dynamics and LO phonon-mediated thermalization in these systems. Here the initial decay time ($\tau_1$) is on the order of 1–5 $ps$, while the second, longer decay ($\tau_2$) occurs over tens of $ps$ (10-120 $ps$) – see Figure 5. The enhanced relaxation time with increasing power density evident for ($\tau_1$) is consistent with the creation of a hot phonon bottleneck [13, 15, 48]. Again, in Figure 5 it is also evident in all cases that the thermalization time is faster under front side illumination (black symbols – via the thin poly-TPD)) as compared to back illumination (red symbols – via the $C_{60}$-ETL). In all cases, the behavior/pattern of the decay times under both front and backside illuminations mirror one other, especially at the highest power, $\sim 20$ W/cm² – see Figures 5(a) and (b). While this trend is less clear for the lower power (15 W/cm²) some consistency is also evident within experimental errors for these data too.

In the case of frontside illumination (black closed squares), the decay time is shorter when current is extracted from the solar cell, namely under $V_{Jsc}$ and $V_{max}$ conditions. In Figure 5, at a bias of 0.02 V (close to or ~$V_{Jsc}$), which approximately represents the maximum current drawn from the solar cell, the lowest decay time is observed, followed by that at a bias of ~ 0.8 V ($V_{max}$) applied to the solar cell, see Figure 1(c). In the case of the non-contacted device conditions, at $V_{oc}$, which represents zero net current extraction from the cell, longer decay times are observed. It therefore appears that carrier thermalization is modulated by carrier extraction (driven by external biasing).

This behavior further supports the relative role of the phonon bottleneck in sustaining (or inhibiting) carrier thermalization in these systems. That is, when photogenerated carriers are rapidly removed from the solar cell in – for example - short circuit conditions the thermalization rate reduces and the carrier dependent phonon bottleneck decreases. If the photogenerated carriers are retained within the device – at open circuit – the relative carrier density is higher and a larger phonon bottleneck results, reducing the relative carrier thermalization.

When considering backside illumination through the ETL (closed red circles – Figure 5), the behavior is somewhat different than that of front side excitation (via HTL), but consistent with earlier work assessing TA measurements of individual HTL/perovskite and ETL/perovskite sub-stacks of the samples studied here, which also showed non-equivalent carrier dynamics in the system [49]. Indeed, despite the relatively cooler $T_c$ with respect to frontside illumination (see Figure 3(b) and (c)), the carrier lifetime under back side illumination is longer under all bias conditions (closed red circles – Figure 5). Moreover, while the hypothesis in terms of the carrier cooling time under front side illumination invokes the relative rates of extraction and inhibited carrier collection at open circuit, $V_{oc}$, (under frontside illumination) and the prevalence of a phonon bottleneck at higher excitation powers (Figure 5(a) and 5(b)): the cooling rate under short circuit, $V_{jsc}$, conditions exceeds that of $V_{oc}$ when the sample is excited from the back via the $C_{60}$ and $SnO_2$ layers. At lower excitation powers (Figure 5(c) and 5(d)) little dependence in cooling rate is seen across the various biasing conditions. While carrier cooling behavior under backside is certainly less easily reconciled in terms of the thermalization properties to those observed under front side excitation, this reflects the relative roles of minority 'hot' electrons (holes) transport and extraction when excited under the different orientations.

Once again, the $T_c$ extracted from the TA (Figure 1(d)) reflects the average temperature of the carriers, comprising both electron and hole carrier thermalization, as such little is known *directly* regarding the relative temperature and thermalization of these constituent carriers. Under backside illumination the photogenerated electrons travel further than the holes prior to collection, while the case is reverse under frontside illumination. It is postulated that the collection of the photogenerated electrons generated at excitation at the backside of the structure are collected less efficiently than that of the electrons and that this increases the relative thermalization rate of hot electrons under backside illumination, despite the lower relative carrier density generated in the absorber under this excitation orientation (Figure 3(b) and (c)). While further work is required to support this hypothesis, it is consistent with the data presented and earlier work that show parasitic barriers in similar metal perovskite solar cells at the perovskite/$SnO_2$-$C_{60}$ interface that require efficient electron tunneling and thermionic emission rate to operate efficiently and are exacerbated at lower temperature [27, 50, 51].

**Conclusion**

In this study, it is shown that the behavior of photogenerated hot carriers in metal halide perovskite solar cells differs significantly when considering their dynamics in non-contacted devices at open circuit, $V_{oc}$, versus and when the cells are operating under at short circuit conditions, $V_{Jsc}$, or at the maximum power point, $V_{max}$ This is attributed to the relative contribution of carrier density, the role of field aided or diffuse transport and/or collection and the respective role of these properties on the creation of a carrier-LO phonon bottleneck particularly in the sub-*ps* regime where strong carrier-phonon dynamics are at play.

**Conflict of Interest Statement**

Authors declare no conflict of interest.

**Data Availability**

Data for this article, including [power and bias dependent TA and J-V results] are available at [SHAREOK repository] at [xmlui.XMLWorkflow.default.def-editstep.claimaction].

**Acknowledgments**

Support is acknowledged from Department of Energy EPSCoR Program and the Office of Basic Energy Sciences, Materials Science, and Energy Division under Award DE-SC0019384. IRS and VRW also acknowledge support from the Center for Advanced Semiconductor Technologies (CAST) at the University of Buffalo. VM and HA are grateful for the support received from the Dodge Family postdoctoral fellowship program at the University of Oklahoma. This material is based in part upon work supported by the National Science Foundation under Grant No. OISE- 2230706.

# Hot Carrier Dynamics in Operational Metal Halide Perovskite Solar Cells

Hadi Afshari[1], Varun Mapara[1], Shashi Sourabh[1], Megh N. Khanal[2], Vincent R. Whiteside[2], Rebecca A. Scheidt[3], Matthew C. Beard[3], Giles E. Eperon[4], Ian R. Sellers[2], and Madalina Furis[1]

[1] *DEPARTMENT OF PHYSICS & ASTRONOMY, UNIVERSITY OF OKLAHOMA, NORMAN, OK 73019, U.S.A*

[2] *DEPARTMENT OF ELECTRICAL ENGINEERING, UNIVERSITY AT BUFFALO, NY 14260, U.S.A*

[3] *NATIONAL RENEWABLE ENERGY LABORATORY, GOLDEN, CO 80401, U.S.A*

[4] *SWIFT SOLAR, SAN CARLOS, CA, 94070, U.S.A*

The carrier temperature, $T_C$ is determined from fitting the high-energy tail of the transient absorption (TA) curves by employing the Fermi-Dirac distribution function. Under hot carrier conditions, where $((E - E_F) \gg k_B T_c)$, this function simplifies to the Maxwell–Boltzmann distribution.

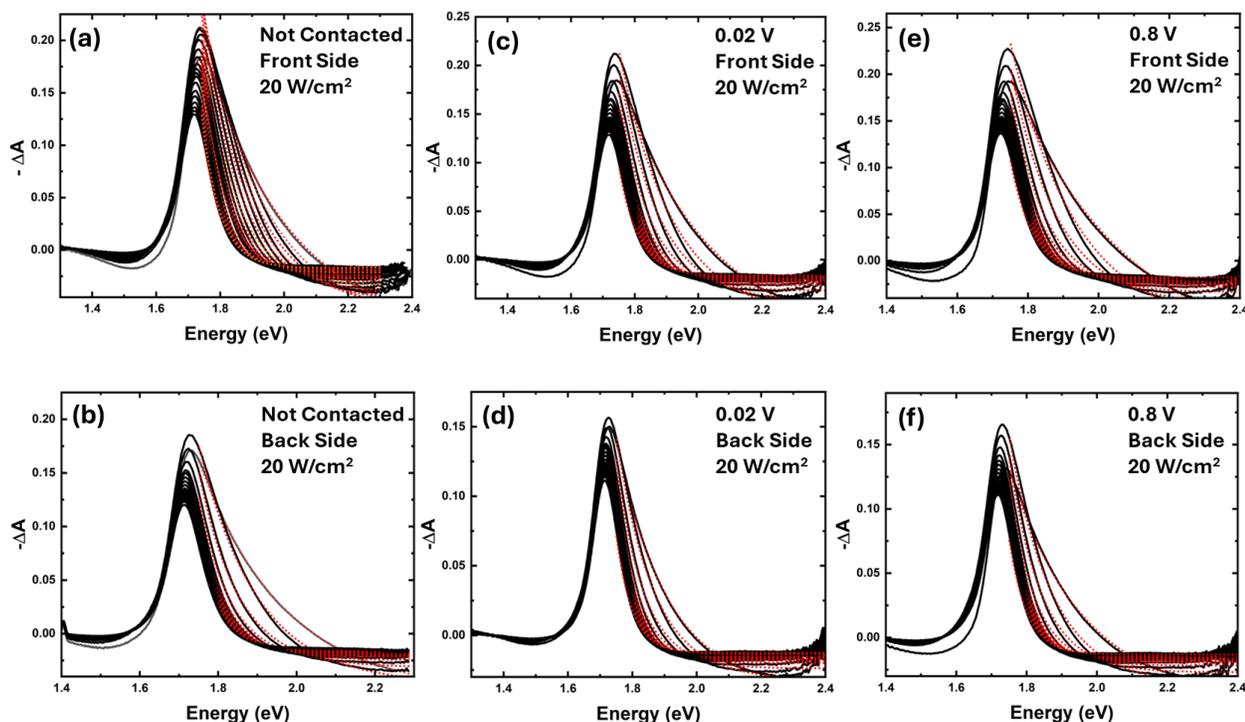

Figure S1. Transient absorption results were recorded at various time delays up to 100 ps. Figures (a) and (b) illustrate the results for the non-contacted ($V_{oc}$) case at a pump power of 20 W/cm² under front-side and back-side illumination, respectively. Figures (c) and (d) show the results for a bias of 0.02 V ($V_{JSC}$), again under front-side and back-side illumination, respectively.

Similarly, figures (e) and (f) present the results for a bias of 0.8 V ($V_{max}$) under front-side and back-side illumination, respectively.

In all measurements, the pump power was maintained at 20 W/cm². The top row of figures corresponds to front-side illumination, while the bottom row represents back-side illumination. As observed, the magnitude of the bleach for back-side illumination (bottom row) is significantly lower compared to front-side illumination. This reflects the lower absorption in the perovskite layer in the case of backside illumination.

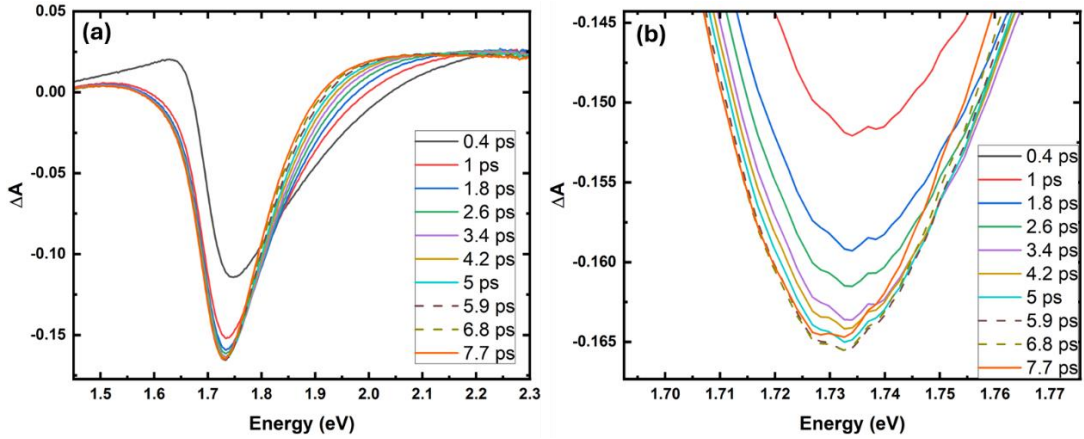

Figure S2. $\Delta A$ as a function of energy at various time delays up to 7.7 ps. Up to 5.9 ps mark the magnitude of bleach increases and above that point it starts reducing systematically for higher delay times.

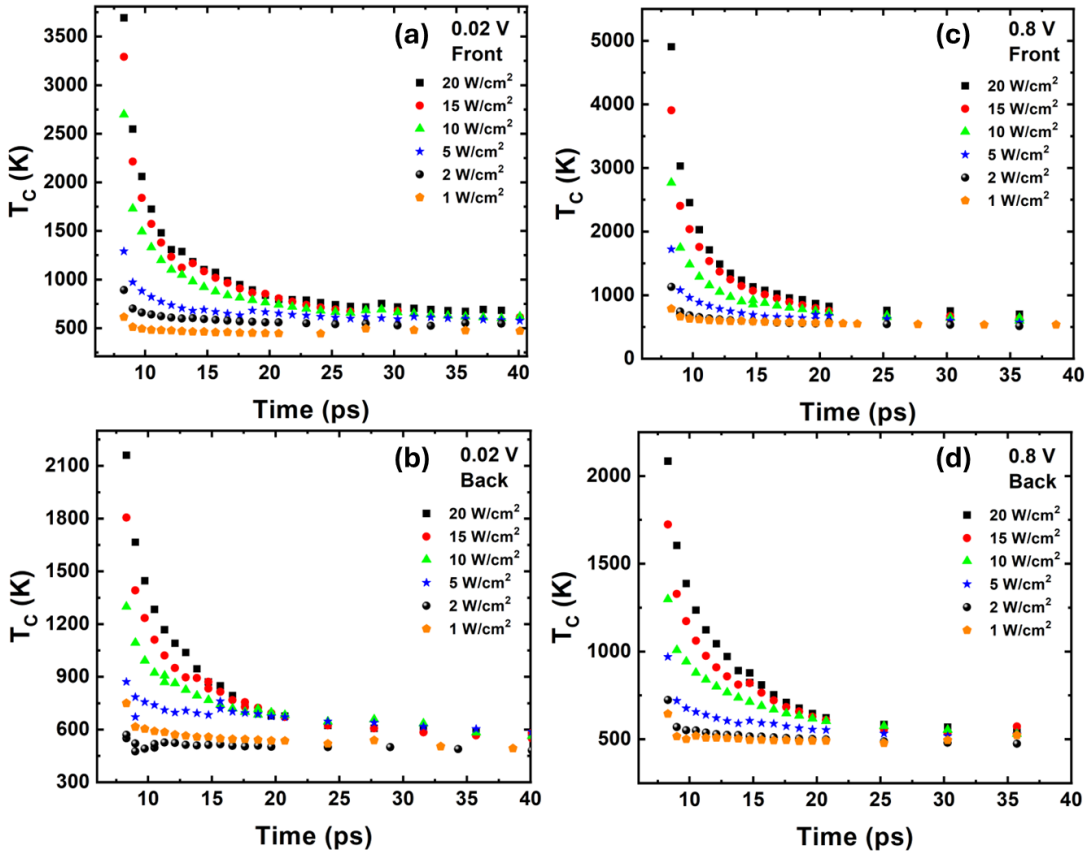

Figure S3. Temperature of hot carriers as a function of time for various biases and pump powers. Figure (a) and (b) show the results for external bias of 0.02 V front and back illumination respectively. Figures (c) and (d) show the results for external bias of 0.8 V front and back illumination respectively. The carrier temperature curves are fitted with biexponential functions to extract the cooling times. If a biexponential fit is not possible one can use triexponential (or even four exponential) functions to acquire meaningful relaxation decay times.

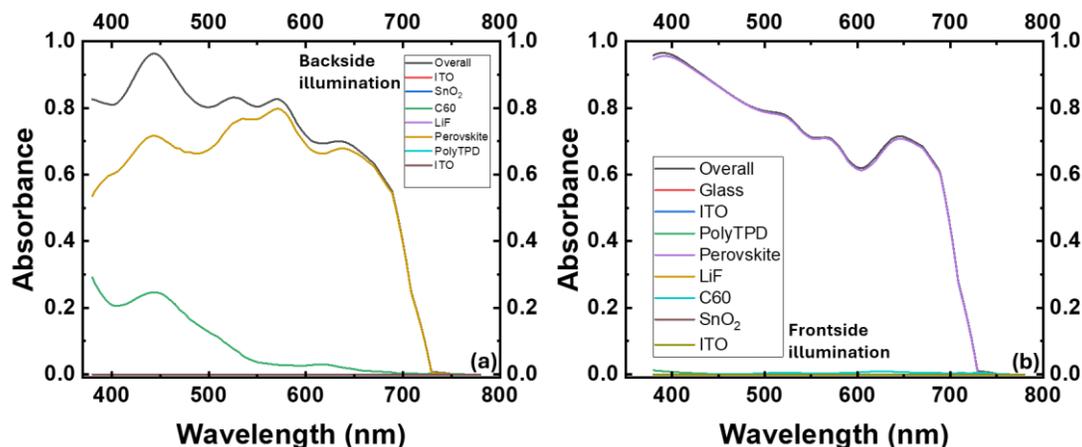

Figure S4. Based on simulations in Setfos software, which utilizes transfer matrix formalism, the layer absorbance profile shows that when the solar cell is illuminated from the back contact side (Figure a), the $C_{60}$ layer absorbs significantly in the visible spectrum. In contrast, when light is incident from the front glass side (Figure b), most of the light is absorbed directly in the perovskite absorber material and the $C_{60}$ layer displays much lower absorption levels.